\documentclass[a4paper,11pt]{article}
\usepackage{jheppub} 
\usepackage[T1]{fontenc} 
\RequirePackage[displaymath]{lineno} % Display line numbers
\usepackage{epsfig}
\usepackage{graphicx}% Include figure files
\usepackage{dcolumn}% Align table columns on decimal point
\usepackage{bm}% bold math
\usepackage{ltablex,booktabs}
\usepackage{overpic}
\usepackage{subfigure}
\usepackage{float}
\usepackage{color}
\usepackage{amsmath}
\usepackage{mathcomp}
\usepackage{mathrsfs}
\usepackage{multirow}
\usepackage{rotating}
\usepackage{amssymb}
\usepackage{gensymb}
\usepackage{amsmath}
\usepackage{tabularx}
\usepackage{epstopdf}
\usepackage{ulem}

\title{Observation of the semileptonic decays $D^0\rightarrow K_S^0\pi^-\pi^0 e^+ \nu_e$ and $D^+\rightarrow K_S^0\pi^+\pi^- e^+ \nu_e$}
\collaborationImg{\includegraphics[width=0.15\textwidth, angle=90]{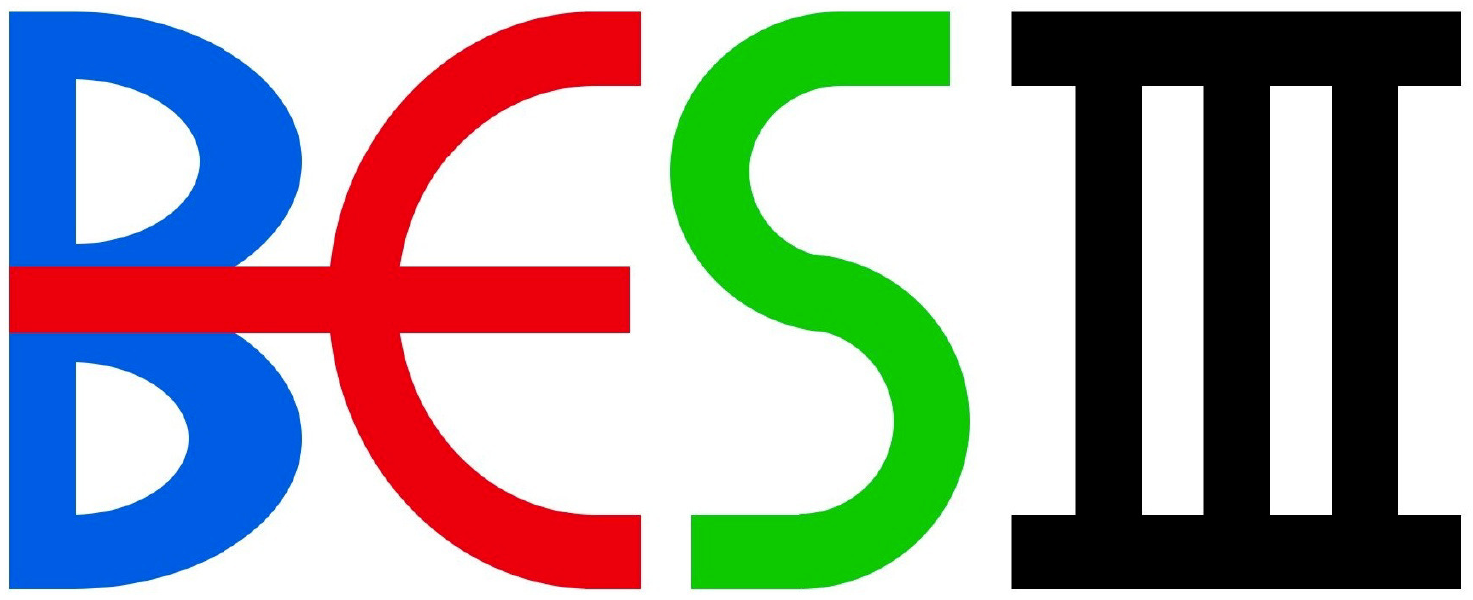}}
\collaboration{The BESIII Collaboration}
\abstract{By analyzing $e^+e^-$ annihilation data corresponding to an integrated luminosity of 2.93 $\rm fb^{-1}$ collected at a center-of-mass energy of 3.773~GeV with the \text{BESIII} detector, the first observation of the semileptonic decays $D^0\rightarrow K_S^0\pi^-\pi^0 e^+ \nu_e$ and $D^+\rightarrow K_S^0\pi^+\pi^- e^+ \nu_e$  is reported. With a dominant hadronic contribution from $K_1(1270)$, the branching fractions are measured to be $\mathcal{B}(D^0\rightarrow {K}_1(1270)^-(\to K^0_S\pi^-\pi^0)e^+\nu_e)=(1.69^{+0.53}_{-0.46}\pm0.15)\times10^{-4}$ and $\mathcal{B}(D^+\to \bar{K}_1(1270)^0(\to K^0_S\pi^+\pi^-)e^+\nu_e)=(1.47^{+0.45}_{-0.40}\pm0.20)\times10^{-4}$ with statistical significance of 5.4$\sigma$ and 5.6$\sigma$, respectively.  When combined  with measurements of the $K_1(1270)\to K^+\pi^-\pi$ decays, the absolute branching fractions are determined to be $\mathcal{B}(D^0\to  K_1(1270)^-e^+\nu_e)=(1.05^{+0.33}_{-0.28}\pm0.12\pm0.12)\times10^{-3}$ and $\mathcal{B}(D^+\to \bar{K}_1(1270)^0e^+\nu_e)=(1.29^{+0.40}_{-0.35}\pm0.18\pm0.15)\times10^{-3}$. The first  and second  uncertainties are statistical and systematic, respectively, and the third  uncertainties originate  from the assumed branching fractions of  the  $K_1(1270)\to K\pi\pi$ decays.
}

\arxivnumber{}
\begin{document}
% \begin{linenumbers}
\maketitle
\flushbottom
\def\Dp{D^{+}}
\def\Dm{D^{-}}
\def\pimp{\pi^{\mp}}
\def\pipm{\pi^{\pm}}
\def\pip{\pi^{+}}
\def\pim{\pi^{-}}
\def\piz{\pi^{0}}
\def\ks{K_{S}^{0}}
\def\ee{e^{+}e^{-}}
\def\dedx{\mathrm{d}E/\mathrm{d}x}
\def\e{\mathrm{e}}
\def\Dk1enu{$D\rightarrow K_1(1270)e^+\nu_e$}
\def\kzero{\ensuremath{K^0}}
\def\kzerobar{\ensuremath{\overline{K}\vphantom{K}^0}}
\def\kaon{K}
\def\kp{K^+}
\def\km{K^-}
\def\kpm{K^{\pm}}
\def\kmp{K^{\mp}}
\def\BR{\mathcal{B}}
\def \gev  {\mbox{GeV}}
\def \gevc {\mbox{GeV/$c$}}
\def \gevcc{\mbox{GeV/$c^2$}}
\def \mev  {\mbox{MeV}}
\def \mevcc{\mbox{MeV/$c^2$}}
\def \ipb  {\mbox{pb$^{-1}$}}
\def \ifb  {\mbox{fb$^{-1}$}}
\def \kpipi{K^{+}\pi^{-}\pi^{-}}
\def \kpipipiz{K^{+}\pi^{-}\pi^{-}\pi^{0}}
\def \kspipiz{\ks\pi^{-}\pi^{0}}
\def \kspi{\ks\pi^{-}}
\def \kspipipi{\ks\pi^{-}\pi^{-}\pi^{+}}
\def \kkpi{K^{+}K^{-}\pi^{-}}
\def \kpipiz{K^{-}\pi^{+}\pi^{0}}
\def \ss {\sqrt{s}}
\def \mbc {M_{\rm{BC}}}
\def \msigmap {M_{p\pi^0}}
\def \mbcst {M^{\rm{ST}}_{\rm{BC}}}
\def \dEst {\Delta E_{\rm{ST}}}
\def \mbctag {M^{\rm{tag}}_{\rm{BC}}}
\def \mbcsig {M^{\rm{sig}}_{\rm{BC}}}
\def \dE {\Delta E}
\def \dEtag {\Delta E_{\rm{tag}}}
\def \dEsig {\Delta E_{\rm{sig}}}
\def \ebeam {E_{\rm{beam}}}
\def \mrec {M_{\rm{rec}}}
\def \mkppiz {M^{2}_{K^{+}\pi^{0}}}
\def \mkspiz {M^{2}_{K_{S}^{0}\pi^{0}}}
\def \mkpks {M^{2}_{K^{+}K_{S}^{0}}}
\def \kstarp {K^{*}(892)^{+}}
\def \kstarz {\bar{K}^*(892)^0}
\def \kppizswave {(K^{+}\pi^{0})_{\mathcal{S}\mathrm{-wave}}}
\def \ksstarp {K^{*}(1410)^{+}}
\def \ksstarz{\bar{K}^{*}(1410)^{0}}
\def \ap {a_{0}(980)^{+}}
\def \app {a_{0}(1450)^{+}}
\def \romanOne   {\uppercase\expandafter{\romannumeral1}}
\def \romanTwo   {\uppercase\expandafter{\romannumeral2}}
\def \romanThree {\uppercase\expandafter{\romannumeral3}}
\def \romanFour  {\uppercase\expandafter{\romannumeral4}}
\def \romanFive  {\uppercase\expandafter{\romannumeral5}}
\def \romanSix   {\uppercase\expandafter{\romannumeral6}}
\def \romanSeven {\uppercase\expandafter{\romannumeral7}}
\def \romanEight {\uppercase\expandafter{\romannumeral8}}
\def \romanNine {\uppercase\expandafter{\romannumeral9}}
\newcommand{\lambdacpm}{\Lambda_{c}^{\pm}}
\newcommand{\lamcplamcm}{\Lambda_{c}^{+}\bar{\Lambda}_{c}^{-}}
\newcommand{\lambdacp}{\Lambda_{c}^{+}}
\newcommand{\lambdacm}{\bar{\Lambda}_{c}^{-}}
\def\kshort{K^0_{\mathrm{S}}}
\newcommand{\sigmode}[1]{
	\ifnum#1=1
	\lambdacp \rightarrow \Sigma^0 K^+
	\else
	\ifnum#1=2
	\lambdacp \rightarrow \Sigma^+ \kshort
	\fi
	\fi
}
\newcommand{\refmode}[1]{
	\ifnum#1=1
	\lambdacp \rightarrow \Sigma^0 \pi^+
	\else
	\ifnum#1=2
	\lambdacp \rightarrow \Sigma^+ \pi^+\pi^-
	\fi
	\fi
}
\newcommand{\sigmodefs}[1]{
	\ifnum#1=1
	\Sigma^0 K^+
	\else
	\ifnum#1=2
	\Sigma^+ \kshort
	\fi
	\fi
}
\newcommand{\refmodefs}[1]{
	\ifnum#1=1
	\Sigma^0 \pi^+
	\else
	\ifnum#1=2
	\Sigma^+ \pi^+\pi^-
	\fi
	\fi
}
\newcommand{\bkgmode}{
	\lambdacp \rightarrow p \ks \piz
}

%%%%%%%%%%%%%%%%%%%%%%%%%%%%%%%%%%%%%%%%%%%%%%%%%%%%%%%%%%%%%%%%%%%%%%%%%%%%%%%%%%%%%%%%%%%%%%%%%%%%

\newcommand{\mdp}{D^{+}}
\newcommand{\mdm}{D^{-}}
\newcommand{\mdz}{D^{0}}
\newcommand{\ddp}{$D^{+}$ }
\newcommand{\ddm}{$D^{-}$ }
%%%%%%%%%%%%%%%%%%%%%%%%%%%%%%%%%%%%%%%%%%%%%%%%%%%%%%%%%%%%%%%%%%%%%%%%%%%%%%%%%%%%%%%%%%%%%%%%%%%%

\newcommand{\mdpdm}{D^{+}D^{-}}
\newcommand{\ddpdm}{$D^{+}D^{-}$ }
\newcommand{\dzdzbar}{$D^{0}\bar{D}^{0}$ }
\newcommand{\dmknpi}{$mKn\pi$ }
\newcommand{\mmknpi}{\rm mKn\pi}
\newcommand{\umiss}{U_{\rm miss}}
\newcommand{\dumiss}{$U_{\rm miss}$ }
%%%%%%%%%%%%%%%%%%%%%%%%%%%%%%%%%%%%%%%%%%%%%%%%%%%%%%%%%%%%%%%%%%%%%%%%%%%%%%%%%%%%%%%%%%%%%%%%%%%%
\makeatletter
\newcommand{\rmnum}[1]{\romannumeral #1}
\newcommand{\Rmnum}[1]{\expandafter\@slowromancap\romannumeral #1@}
\makeatother
%%%%%%%%%%%%%%%%%%%%%%%%%%%%%%%%%%%%%%%%%%%%%%%%%%%%%%%%%%%%%%%%%%%%%%%%%%%%%%%%%%%%%%%%%%%%%%%%%%%%
\newcommand{\mkso}{K^{0}_{S}}
\newcommand{\modefir}{K^{+}\pi^{-}\pi^{-}}
\newcommand{\modesec}{K^{+}\pi^{-}\pi^{-}\piz}
\newcommand{\modethi}{\mkso\pi^{-}}
\newcommand{\modefou}{K^{+}K^{-}\pi^{-}}
\newcommand{\modefif}{\mkso\pi^{+}\pi^{-}\pi^{-}}
\newcommand{\modesix}{\mkso\pi^{-}\piz}
\newcommand{\ddmodefir}{$K^{+}\pi^{-}\pi^{-}$}
\newcommand{\ddmodesec}{$K^{+}\pi^{-}\pi^{-}\piz$}
\newcommand{\ddmodethi}{$\mkso\pi^{-}$}
\newcommand{\ddmodefou}{$K^{+}K^{-}\pi^{-}$}
\newcommand{\ddmodefif}{$\mkso\pi^{+}\pi^{-}\pi^{-}$}
\newcommand{\ddmodesix}{$\mkso\pi^{-}\piz$}
%%%%%%%%%%%%%%%%%%%%%%%%%%%%%%%%%%%%%%%%%%%%%%%%%%%%%%%%%%%%%%%%%%%%%%%%%%%%%%%%%%%%%%%%%%%%%%%%%%%%

\newcommand{\zksoenue}{D^{0} \rightarrow {K}^{-}e^{+}\nu_{e}}
\newcommand{\dksoenue}{$D^{+} \rightarrow {\bar{K}}^{0}e^{+}\nu_{e}$}
\newcommand{\dzksoenue}{$D^{0} \rightarrow {K}^{-}e^{+}\nu_{e}$}
\newcommand{\tagsemimc}{``Cocktail VS Cocktail $D\bar{D}$ decay'' Monte Carlo }

\newcommand{\koneev}{$D^{+} \rightarrow {\bar{K}}^{0}_{1}(1400)e^{+}\nu_{e}$}
\newcommand{\kdecay}{$D^{+} \rightarrow {\bar{K}}^{0}_{1}(1270)e^{+}\nu_{e}$}
\newcommand{\Ddecaydecay}{$$D^+\to K_S^0\pi^+\pi^-e^+\nu_e$}
\newcommand{\dokoneenue}{$D^{0} \rightarrow K^{0}_{1}(1270)e^{+}\nu_{e}$}
\newcommand{\bkoneenue}{${\mathcal B} (D^{+} \rightarrow {\bar{K}}^{0}_{1}(1270)e^{+}\nu_{e})$}
\newcommand{\bkoneev}{${\mathcal B} (D^{+} \rightarrow {\bar{K}}^{0}_{1}(1400)e^{+}\nu_{e})$}
\newcommand{\kone}{${\bar{K}}^{0}_{1}(1270) \rightarrow K_S^{0} \pi^{+} \pi^{-}$}
\newcommand{\bkone}{${\mathcal B} ({\bar{K}}^{0}_{1}(1270) \rightarrow K^{-} \pi^{+} \pi^{0})$}
%\newcommand{\kone}{${\mathcal B} ({\bar{K}}^{0}_{1}(1270) \rightarrow K^{-} \pi^{+} \pi^{0})$}

%\normalsize
%\parskip=5pt plus 1pt minus 1pt

\section{Introduction}
\label{int}

    Semileptonic charm decays induced by the $c\to s e^+\nu_e$ process are dominated by pseudoscalar ($K$) and vector ($K^*(892)$) mesons,  i.e. contain a kaon and at most one pion in the final-state hadronic systems~\cite{lepdecay,isgw2}. Semileptonic decays with higher multiplicity final states  involving  a kaon and two pions are highly suppressed and are expected to be mostly mediated by the axial-kaon system with a mixing angle $\theta_{K_1}$~\cite{mixing}. Thus knowledge of $\theta_{K_1}$ is essential for theoretical calculations describing the decays of $D$ particles into strange axial-vector mesons~\cite{Cheng:2017pcq,Momeni:2020zrb,Momeni:2019uag}. The $D\to K\pi\pi e^+\nu_e$ decays  provide a unique opportunity to study $K_1(1270)$ and $K_1(1400)$ mesons in a clean environment, without any additional hadrons in the final states. Such studies can lead to a better determination  of $\theta_{K_1}$ as well as of the masses and widths of the $K_1$ mesons, which currently all have large uncertainties~\cite{pdg}.  By exploiting the measured properties  of $D\to \bar{K}_1(1270) \ell^+\nu_\ell$ and $B\to K_1(1270)\gamma$ decays,   the photon polarization in $b\to s\gamma$ can be determined  without considerable theoretical ambiguity, according to Refs.~\cite{Wang:2019wee,Bian:2021gwf}.

    The  BESIII  collaboration, performing studies of the hadronic systems  $K^-\pi^+\pi^-$ and $K^-\pi^+\pi^0$,  
 reported the first observation of semileptonic $D$ decays involving a $K_1(1270)$~\cite{liuk,fylan}, and measured the branching fractions (BFs) $\mathcal{B}({D^0\to K_1(1270)^-e^+\nu_e)} = (1.09\pm0.13^{+0.09}_{-0.13}\pm0.12)\times10^{-3}$ and $\mathcal{B}({D^+\to \bar{K}_1(1270)^0e^+\nu_e)} =( 2.30\pm0.26^{+0.18}_{-0.21}\pm0.25)\times10^{-3}$. Here the first  and second uncertainties are statistical and systematic, respectively, and the third uncertainties originate from the assumed BFs of   $K_1(1270)^{0,+}\to K^+\pi^-\pi^{0,+}$~\cite{pdg}.  
 The decays $D^0\to K_S^0\pi^-\pi^0 e^+\nu_e$ and  $D^+\to K_S^0\pi^+\pi^- e^+\nu_e$ have not yet been observed.  In 2011, based on the $K^+\pi^+\pi^-$ system in the decay of $B^+\to J/\psi K^+\pi^+\pi^-$, the Belle collaboration found the   BFs of  $K_1(1270)\to K\rho, K\omega$, and  $K^*(892)\pi$  to be consistent with previous measurements, but reported the measured   BF of $K_1(1270)\to K_0^*(1430)\pi$  to be significantly smaller~\cite{Belle2011}. Furthermore, measurements of the  BF ratio   $R_{K_1(1270)}=\frac{\mathcal{B}_{K_1(1270) \rightarrow K^* \pi}}{\mathcal{B}_{K_1(1270) \rightarrow K \rho}}$ yield different results depending on the decay channels used~\cite{dskkpipi0, 2012D0kkpipi1,kpipipi, lhcb17, 4pi}, whereas they are expected to be identical under the narrow width approximation for the $K_1(1270)$ meson  assuming $C\! P$ conservation in strong decays~\cite{Ddecays}. 
Measurements of the BFs of  ${D\to K_1(1270)(\to K_S^0\pi\pi)e^+\nu_e}$ decays are desirable, as they would  improve  the knowledge of the relative decay rates of $K_1(1270)$ into final states with one kaon and two pions.

    This paper presents the first observation of  the semileptonic decays $D^0\rightarrow K_S^0\pi^-\pi^0 e^+ \nu_e$ and $D^+\rightarrow K_S^0\pi^+\pi^- e^+ \nu_e$ and   measurements of the BFs  of $D^0\to K_1(1270)^- e^+\nu_e$ and $D^+\to \bar{K}_1(1270)^0e^+\nu_e$.  The analysed data samples come from $e^+e^-$ collisions at a center-of-mass energy of  3.773~GeV, which were collected by the BESIII detector operating at the BEPCII storage ring.  These samples correspond  to an integrated luminosity of  $2.93$~$\rm fb^{-1}$ accumulated at the $\psi(3770)$ resonance~\cite{Ablikim:2013ntc}. Throughout this paper, charge conjugate channels are always implied.

\section{Detector and data sets}
\label{sec:detector_dataset}

The BESIII detector~\cite{Ablikim:2009aa} records  $e^+e^-$ collisions provided by the BEPCII storage ring~\cite{BEPCII} in the center-of-mass energy range from 2.0 to 4.95~GeV, with a peak luminosity of $1 \times 10^{33}\;\text{cm}^{-2}\text{s}^{-1}$ achieved at $\sqrt{s} = 3.773\;\text{GeV}$. 
BESIII has collected large data samples in this energy region~\cite{BESIII:2020nme}. The cylindrical core of the BESIII detector covers 93\% of the full solid angle and consists of a helium-based multilayer drift chamber~(MDC), a plastic scintillator time-of-flight system~(TOF), and a CsI(Tl) electromagnetic calorimeter~(EMC), which are all enclosed in a superconducting solenoidal magnet providing a 1.0~T magnetic field. The solenoid is supported by an octagonal flux-return yoke with resistive-plate-counter muon-identification modules interleaved with steel. 
The charged-particle momentum resolution at $1~{\rm GeV}/c$ is $0.5\%$, and the  ${\rm d}E/{\rm d}x$ resolution is $6\%$ for electrons from Bhabha scattering. The EMC measures photon energies with a resolution of $2.5\%$ ($5\%$) at $1$~GeV in the barrel (end-cap) region. The time resolution in the TOF barrel region is 68~ps, while that in the end-cap region is 110~ps. Details about the design and performance of the BESIII detector are given in Ref.~\cite{Ablikim:2009aa}.

Monte Carlo (MC) simulated data samples produced with a {\sc geant4}-based~\cite{geant4} software package, which includes the geometric description of the BESIII detector and the detector response~\cite{detvis}, are used to determine detection efficiencies and to estimate background contributions. The simulation models the beam-energy spread and initial-state radiation (ISR) in the $e^+e^-$ annihilation with the generator {\sc kkmc}~\cite{kkmc,kkmc2}. An `inclusive' MC event sample includes the production of $D\bar{D}$ pairs (including quantum coherence for the neutral $D$ channels), the non-$D\bar{D}$ decays of the $\psi(3770)$, the ISR production of the $J/\psi$ and $\psi(3686)$ states, and the continuum processes incorporated in {\sc kkmc}~\cite{kkmc,kkmc2}.
All particle decays are modeled with {\sc evtgen}~\cite{evtgen,evtgen2} using  BFs either taken from the Particle Data Group (PDG)~\cite{pdg}, when available, or otherwise estimated with {\sc lundcharm}~\cite{lundcharm, lundcharm2}. Final-state radiation~(FSR) from charged final-state particles is incorporated using the {\sc photos} package~\cite{photos}. The total size of the inclusive MC  samples is approximately 10 times that of the data.

The  $D\rightarrow K_1(1270)e^+\nu_e$  decays are simulated with the ISGW2 model~\cite{isgw2}, and the $K_1(1270)$ is allowed to decay  through  all intermediate processes leading to the final state $K_S^0\pi\pi$. The $K_1(1270)$ resonance shape is parameterized by a relativistic Breit-Wigner function with a mass of $(1.253\pm0.007)\;\rm{GeV}/c^2$ and a width of $(90\pm20)\;\rm{MeV}$~\cite{pdg}.  Using the  BFs of $K_1(1270)$  measured by  Belle~\cite{Belle2011}  as input in the simulation gives  good data/MC agreement  in the kinematic distributions~\cite{fylan}. The $e^+e^- \to D\bar D$ signal MC samples, in which the $D$ decays exclusively into signal modes while the $\bar D$ decays inclusively, are used to determine the detection efficiencies.

\section{Measurement method and single-tag  selection}
\label{ST-selection}
    At $\sqrt{s} = 3.773\;\text{GeV}$, the $\psi(3770)$ resonance is produced in electron-positron annihilation, and then decays predominately into $D\bar D$ pairs without accompanying hadron(s), thereby offering a clean environment to investigate  $D$ decays with the double-tag (DT) method~\cite{MARK-III:1985hbd,Li:2021iwf}. In these cases, when a $\bar D$ meson is fully reconstructed, all of the remaining tracks and photons in the event must originate from the accompanying $D$ meson. The fully reconstructed meson is called a  single-tag (ST)  $\bar D$. The ST $\bar D$ mesons are selected by reconstructing a $\bar D^0$ or $D^-$ in one of the following decay modes: $K^+\pi^-$, $K^+\pi^-\pi^0$, $K^+\pi^-\pi^+\pi^-$, $K^+\pi^-\pi^+\pi^-\pi^0$  for neutral tags, and $K^+\pi^-\pi^-$, $K_S^0\pi^-$, $K^+\pi^-\pi^-\pi^0$, $K^0_S\pi^-\pi^0$, $K^+K^-\pi^-$ and $K_S^0\pi^+\pi^-\pi^-$ for charged tags.  Using the ST $\bar D$  samples, the decays of $D\to K_S^0\pi\pi e^+\nu_e$ can be reliably identified from the recoiling tracks as DT events. The BF of the signal decay is then determined by 
% \label{eq}
\begin{equation}
    \mathcal{B}_{\rm sig} = \frac{N_{\rm DT}}{N_{\rm ST}^{\rm tot}\cdot \epsilon_{\rm sig}}, \label{equa}
\end{equation}
    where  $N_{\mathrm{ST}}^{\mathrm{tot}}$ and  $N_{\mathrm{DT}}$ are the ST and DT yields,  $\epsilon_{\text {sig }}=\sum_i\left[\left(\varepsilon_{\mathrm{DT}}^i N_{\mathrm{ST}}^i\right) /\left(\varepsilon_{\mathrm{ST}}^i N_{\mathrm{ST}}^{\mathrm{tot}}\right)\right]$ is the efficiency of detecting the SL decay in the presence of the ST $\bar{D}$ meson, reconstructed in any of the tag modes. Here, $i$ denotes the tag mode, and $\epsilon_{\rm ST}$ and $\epsilon_{\rm DT}$ are the ST and DT efficiencies of selecting the ST and DT candidates, respectively.  Using the BF of  $K_1(1270)\to K_S^0\pi\pi$ given in the PDG~\cite{pdg}, the BFs of the  ${D\to K_1(1270)e^+\nu_e}$ decays can be obtained.
    
    % tracks and PID selection
    For the reconstruction and identification of $K_S^0$, $K^\pm$, $\pi^\pm$ and $\pi^0$,  the same criteria   are used as in Refs.~\cite{BESIII:2016gbw,k0enu,pimunu, a0enu,kmunu,f0enu}. For any selected charged  track, except for those used for reconstructing $K_S^0$ decays, the polar angle $\theta$ with respect to the  $z$-axis (defined as the symmetry axis of the MDC) is required to satisfy $|\!\cos\theta|<0.93$, and the point of closest approach  to the interaction  point (IP) must be within 1 cm in the  plane perpendicular to the $z$ axis and within $\pm10$~$\rm{cm}$ along the $z$ axis.  Particle identification~(PID) for charged tracks combines measurements of the energy deposited in the MDC~(d$E$/d$x$) and the flight time measured in the TOF to form likelihoods $\mathcal{L}(h)~(h=K,\pi)$ for each hadron hypothesis $h$.
Charged kaons and pions are identified by comparing the likelihoods for the kaon and pion hypotheses, $\mathcal{L}(K)>\mathcal{L}(\pi)$ and $\mathcal{L}(\pi)>\mathcal{L}(K)$, respectively. 
    
    % ks selection
    The $K_S^0$  candidates are selected via the  $K_S^0\to \pi^+\pi^-$ decays,  and hence they  are reconstructed from pairs of oppositely charged tracks. For these two tracks, the distance of closest approach to the IP is required to be less than 20~cm along the $z$ axis. The two charged tracks are constrained to originate from a common vertex that is required to be displaced from the IP by a flight distance of at least twice the vertex resolution. The invariant mass of the $\pi^+\pi^-$ pair is required to be within $(0.486, 0.510)\;\text{GeV}/c^2$.
    
    % pi0 selection
     The $\pi^0$  candidates are reconstructed via $\pi^0\to \gamma\gamma$ decays.  Photon candidates  are reconstructed from isolated electromagnetic showers detected in the EMC crystals. The deposited energy is required to be greater than $25$~$(50)$~$\rm{MeV}$ in the barrel~(end-cap) region. To exclude showers that originate from charged tracks, the angle subtended by the EMC shower and the position of the closest charged track at the EMC must be greater than 10 degrees as measured from the IP.  To further suppress fake photon candidates due to electronic noise or beam-related background, the measured EMC time is required to be within  [0, 700] ns from the event start time. The invariant mass of a photon pair is required to be within $(0.115, 0.150)~\rm{GeV}$$/c^2$. To further improve the resolution of $\pi^0$ momentum $\vec{p}_{\pi^0}$, the invariant mass of the photon pair is constrained to the known $\pi^0$ mass~\cite{pdg} by applying a  kinematic fit.

    For the ST candidates $\bar{D}^0 \rightarrow K^{+} \pi^{-}$,  the  background contributions  from cosmic rays and Bhabha events are rejected by using the analogue requirements as described in Ref.~\cite{kpicosmic}. First, the two charged tracks used must have a TOF time difference less than 5 ns and they must not be consistent with being a muon pair or an electron-positron pair. Second, there must be at least one EMC shower with an energy larger than 50 MeV or at least one additional charged track detected in the MDC.

    The tagged $\bar D$ mesons are selected using two variables, the energy difference
\begin{equation}
     \Delta E = E_{\bar D}-E_{\rm beam} 
\end{equation}
    and the  beam-constrained (BC) mass 
\begin{equation}
    M_{\rm BC} = \sqrt{E_{\rm beam}^2/c^4-|\vec p_{\bar D}|^2/c^2},
\end{equation}
    where $E_{\rm beam}$ is the beam energy, and $\vec p_{\bar D}$ and $E_{\bar D}$ are the momentum and the energy of the $\bar D$ candidate in the $e^+e^-$ rest frame. For each tag mode, if there are multiple combinations in an event, only the one giving the minimum $|\Delta E|$ is retained for further analyses. Combinatorial   background contributions are suppressed with a requirement on $\Delta E$ for each tag mode as described in Refs.~\cite{fylan, liuk};  the $\Delta E$ requirements are summarized in  Table~\ref{delEmBC}.

\begin{table}[htb]
    \centering
    \setlength\tabcolsep{7.5pt}
    \def\arraystretch{1.15}
    \caption{Summary of the $\Delta E$ requirements and $M_{\rm BC}$ mass windows for the ten tag modes.}
    \label{delEmBC}
    \begin{tabular}{lcc}\hline
    \bottomrule
    Tag mode & $\Delta E$ (GeV) & $M_{\rm BC}$ $(\textup{GeV}/c^2)$\\
    \hline
    $\bar D^0\to K^+\pi^-$           &(-0.029,0.027) & (1.858,1.874) \\
    $\bar D^0\to K^+\pi^-\pi^0$      &(-0.069,0.038) & (1.858,1.874) \\
    $\bar D^0\to K^+\pi^-\pi^+\pi^-$ &(-0.031,0.028) & (1.858,1.874) \\
    $\bar D^0\to K^+\pi^-\pi^+\pi^-\pi^0$ &(-0.040,0.025) & (1.858,1.874) \\
    \hline
    $D^-\to\modefir$                & (-0.025,0.025) & (1.863,1.877) \\
    $D^-\to\modesec$                & (-0.055,0.040)  & (1.863,1.877) \\
    $D^-\to\modethi$                & (-0.025,0.025) & (1.863,1.877)\\
    $D^-\to\modesix$                & (-0.055,0.040)  & (1.863,1.877) \\
    $D^-\to\modefif$                & (-0.025,0.025) & (1.863,1.877) \\
    $D^-\to\modefou$                & (-0.025,0.025) & (1.863,1.877) \\
    \bottomrule
    \end{tabular}
\end{table}

    To extract the  yields of ST $\bar D$ mesons for each tag mode, binned maximum-likelihood fits are performed to the $M_{\rm BC}$ distributions of the accepted ST candidates. The signal is modeled by the MC-simulated shape convolved with a double-Gaussian function. The combinatorial background shape is described by an ARGUS function~\cite{ARGUS}. All parameters of the double-Gaussian function and the ARGUS function are  left free in the fit. The numbers of ST $\bar D$ mesons are obtained by integrating over the $\bar D$ signal shape in the mass windows~\cite{fylan, liuk}  which are listed in Table~\ref{delEmBC}. The $M_{\rm BC}$ distributions of the accepted ST candidates in data for the ten tag modes are shown in Figure~\ref{data_mBC}. The ST yield and the ST  efficiency ($\epsilon_{\rm ST}^i$) for each tag mode are summarized in Table~\ref{signal_eff}. The total ST yields of $\bar D^0$ and $D^-$ candidates are $(250.5\pm 0.4_{\rm stat.})\times 10^{4}$ and $(153.2\pm0.3_{\rm stat.})\times 10^4$, respectively.
\begin{figure}[h]
    \centering
    \includegraphics[width=5in]{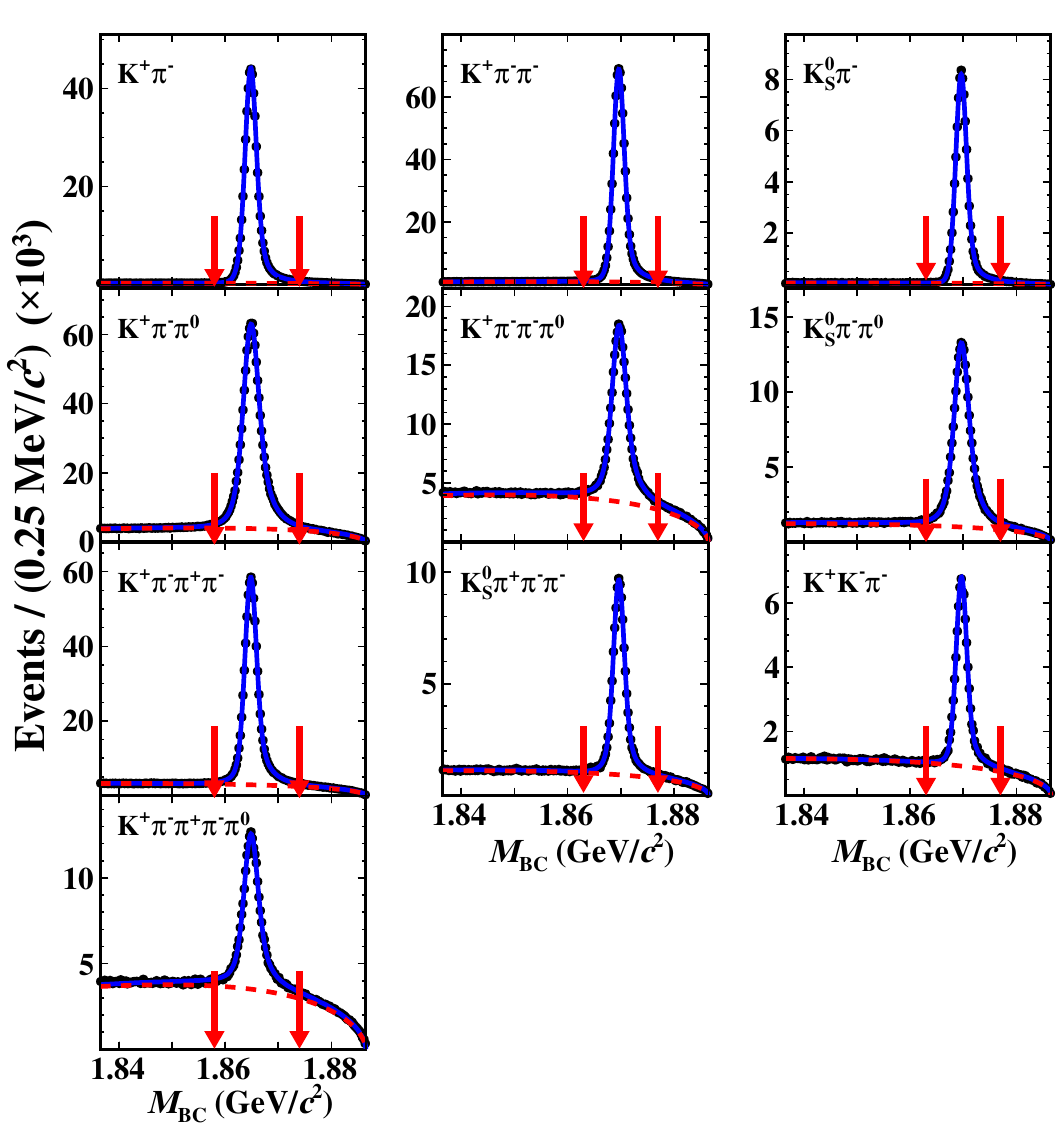}
    \caption{Fits to the $M_{\rm BC}$ distributions of the ST $\bar D$ candidates. The dots with error bars are the real data, the red dashed lines denote the background and the blue solid lines represent the overall fit.  The arrows indicate the limits of  the $M_{\rm BC}$ signal window. }
    \label{data_mBC}
\end{figure}

\section{Branching fractions}
\subsection{Selection of signal candidates}

%% DT Reconstruction
Candidates for the semileptonic decays $D\to K_S^0\pi\pi e^+\nu_e$ are reconstructed from the remaining tracks and showers that have not been used for the ST $\bar D$ reconstruction. The $K_S^0$, $\pi^\pm$ and  $\pi^0$ candidates are selected with the same criteria as on the tag side. Positron PID uses the measured information in the MDC, TOF and EMC. Combined likelihoods ($\mathcal{L}'$) are calculated under the positron, pion, and kaon hypotheses. Positron candidates are required to satisfy $\mathcal{L}'(e)>0.001$ and $\mathcal{L}'(e)/(\mathcal{L}'(e)+\mathcal{L}'(\pi)+\mathcal{L}'(K))>0.8$. To reduce background from hadrons, the positron candidate is further required to have a deposited energy in the EMC and momentum which satisfy   $E/|\vec{p}|c >0.18$   $\times\chi^2_{\textup{d}E/\textup{d}x}+0.32$~\cite{fylan}, where $E$ and $\vec p$ are the energy and  momentum  of positrons, $\chi^2_{\textup{d}E/\textup{d}x}$ is the  difference between the measured energy loss and the expectation from the Bethe-Bloch curve normalized by the resolution for positrons. 
To partially compensate for the energy loss 
due to FSR and bremsstrahlung, the four momenta of neighboring photons with an energy greater than 50~MeV and within a cone of 5 degrees around the positron direction,  are added back to the four-momenta of the positron candidates (FSR  recovery).

When reconstructing the $D^0\rightarrow K_S^0\pi^-\pi^0e^+\nu_e$ decay, the $\pi^0$ mesons  are required to have an energy greater than 0.22~GeV and a decay angle  $\theta_{\pi^0}$  defined through $|\cos\theta_{\pi^0}| = |E_{\gamma_1} -E_{\gamma_2}|/|\vec{p}_{\pi^0}|c$, less than 0.83 to effectively veto fake $\pi^0$ candidates. Here, $E_{\gamma_1}$ and $E_{\gamma_2}$ are the energies of the two daughter photons of  the $\pi^0$ candidate, and $\vec{p}_{\pi^0}$ is its reconstructed momentum. 
To suppress the background from $D^0\to K^0_S\pi^+\pi^-\pi^0$ decays,  $M_{K^0_S\pi^-\pi^0\pi^+_{e\to\pi}}<1.78\;\text{GeV}/c^2$  is required, where $\pi^+_{e\to\pi}$ is the positron candidate reconstructed under the pion mass hypothesis.

For  the  $D^+\rightarrow K_S^0\pi^+\pi^-e^+\nu_e$ decay, the positron must have the opposite charge to that of the tagged $D^-$ meson and the two charged pions must have opposite charge. To suppress the background from $D^+\to K_S^0\pi^+\pi^-\pi^+$ decays, the mass $M_{K_{S}^{0}\pi^{+}\pi^{-}\pi^{+}_{e\to\pi}}$ is required to be less than 1.83 $\rm{GeV}/$$c^2$. In order to reject background events from $D^+\to K_S^0\pi^+\pi^0$ with the $\pi^0$ Dalitz decay $\pi^0\to e^+e^-\gamma$, the opening angle $\theta_\alpha$ between $e^+$ and  $\pi^-$ is required to satisfy $\cos\theta_\alpha<0.95$. To reject contamination from $D^+\to K_S^0\pi^+\pi^-\pi^+\pi^0$ decays, the mass $M_{K_S^0\pi^+\pi^-\pi^+_{e\to \pi}\pi^0}$ is required to satisfy $M_{K_S^0\pi^+\pi^-\pi^+_{e\to \pi}\pi^0}<$1.4~GeV$/c^2$ when there is at least one $\pi^0$ candidate  recoiling against the ST  $D^-$ meson in the event. Furthermore, the opening angle $\theta_\beta$ between the missing momentum (defined below) and the most energetic unused   shower  is required to satisfy $\cos\theta_\beta<0.88$. 

\subsection{Measurement of branching fractions}

    To obtain information about the undetected neutrino, a kinematic quantity is defined as
\begin{equation}
	M_{\rm miss}^2 = E_{\rm miss}^2/c^4 - |\vec{p}_{\rm miss}|^2/c^2,
\end{equation}
where $E_{\rm miss}$ and $\vec{p}_{\rm miss}$ are the total energy and momentum of all missing particles in the event, respectively. They are calculated using $E_{\rm miss}=E_{\rm beam}-\sum_i E_i$, $\vec p_{\rm miss} = -\vec{p}_{D}-\sum_i \vec p_i$ where  $E_i$ and $\vec p_i$ are the measured energy and momentum of particle $i$ in the $e^+e^-$ center-of-mass frame, and $i$ runs over $K_S^0$, $\pi^\pm$, $\pi^0$ and $e^+$ of the signal candidate. 
     In order to improve the $M_{\rm miss}^2$ resolution,  a four-constraint (4-C) kinematic fit is employed.  Here energy and momentum conservation is imposed, and the invariant masses of the $D^0$  and $D^+$ candidate particles are constrained to their known values.
Then the momenta and energies from the kinematic fit are used to calculate $M^2_{\rm{miss}}$.

\begin{figure}[htb]
    \centering
    \includegraphics[scale=0.8]{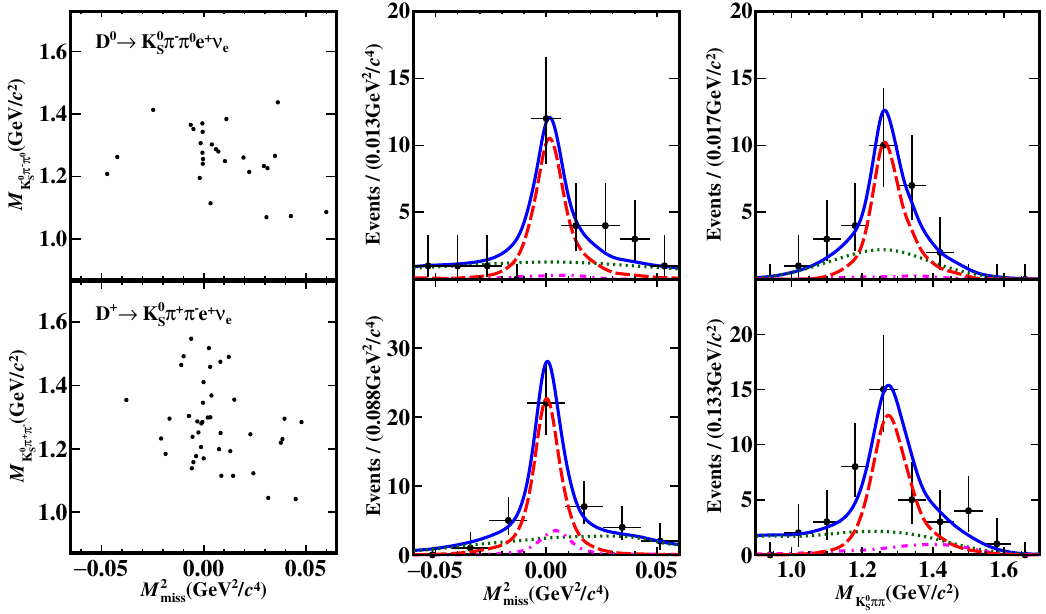}
    \caption{Distributions of $M_{\rm miss}^2$  versus $M_{K_S^0\pi\pi}$ for the accepted semileptonic candidates 
 (left column) and the projections on $M^2_{\rm miss}$  (middle column)  and $M_{K_S^0\pi\pi}$ (right column) of the two-dimensional fits.  The top row is for $D^0\rightarrow K_S^0\pi^-\pi^0e^+\nu_e$ and the bottom row is for $D^+\rightarrow K_S^0\pi^+\pi^-e^+\nu_e$. The dots with error bars are 
 data. The blue solid line denotes the total fit. The red  dashed line represents the signal.  The green  dotted  and purple  dash dotted lines represent the combinatorial background  and  peaking background of $D\to K_S^0 \pi\pi\pi$, respectively.}
    \label{combfit}
\end{figure}

The distributions of $M^2_{\rm{miss}}$ versus $M_{K_S^0\pi\pi}$ distributions of the candidate events  for $D^0\to K_S^0\pi^-\pi^0e^+\nu_e$ and $D^+\to K_S^0\pi^+\pi^-e^+\nu_e$  surviving in data are shown in Figure~\ref{combfit} after combining all tag modes for $D^0$  or $D^+$. Signal events concentrate around zero in $M_{\rm{miss}}^2$.  They are found to cluster around the $K_1(1270)$ nominal mass in the $M_{K_S^0\pi\pi}$ distribution. To determine the signal yield, a two-dimensional unbinned extended maximum-likelihood fit is performed on the $M_{\rm{miss}}^2$ versus $M_{K_S^0\pi\pi}$ distributions of the accepted $D^0\to K_S^0\pi^-\pi^0e^+\nu_e$ and $D^+\to K_S^0\pi^+\pi^-e^+\nu_e$ candidate events, respectively. The two-dimensional signal  and background shapes are derived from  the signal  and inclusive MC samples, respectively. The numbers of peaking background events are fixed  based on the estimation from the simulated samples, while the yields of signal and combinatorial background are free parameters. The two-dimensional probability density functions of signal and background are modeled by using RooNDKeysPdf~\cite{roofit}.

\begin{table}[htb]
    \centering
    \setlength\tabcolsep{7pt}
    \caption{Summary of ST yields $N_{\rm ST}^i$, ST efficiencies $\epsilon_{\rm ST}^i$~$(\%)$,  signal efficiencies $\epsilon_{\rm sig}^i$ of different tag modes $i$, where the uncertainties are statistical. The last  two   columns are the weighted efficiencies $\bar{\epsilon}_{\rm sig}$  and signal yields $N_{\rm DT}$.}
    \label{signal_eff}
    % \bgroup
    \def\arraystretch{1.15}%  1 is the default, change whatever you need
    \begin{tabular}{lrcccc}
      \hline
      \bottomrule
      Tag mode & $N_{\rm ST}^i$~$(\rm \times10^3)$ & $\epsilon_{\rm ST}^i$~$(\%)$ & $\epsilon_{\rm sig}^i$~$(\%)$& $\bar{\epsilon}_{\rm sig}$~$(\%)$ & $N_{\rm DT}$\\\hline
      $\bar{D}^0\to K^+\pi^-$    &540.7$\pm$0.8  & $66.57\pm0.09$ &  4.86$\pm$0.05&
      \multirow{4}{*}{3.89$\pm$0.03}&\multirow{4}{*}{$16.5^{+5.1}_{-4.5}$ }\\
      $\bar{D}^0\to K^+\pi^-\pi^0$             &1066.8$\pm$1.2  & $34.76\pm0.04$ &4.10$\pm$0.03\\
      $\bar{D}^0\to K^+\pi^-\pi^+\pi^-$        &736.3$\pm$1.2   & $41.22\pm0.05$  &3.30$\pm$0.04\\
      $\bar{D}^0\to K^+\pi^-\pi^+\pi^-\pi^0$   &162.0$\pm$0.4   & $15.94\pm0.05$  &2.02$\pm$0.04 \\

      \hline
      $D^-\to K^{+}\pi^{-}\pi^{-}$             &800.6$\pm$1.0   & $51.10\pm0.06$  &9.99$\pm$0.11&
      \multirow{6}{*}{8.96$\pm$0.09}& \multirow{6}{*}{$20.2^{+6.2}_{-5.4}$}\\
      $D^-\to K^{+}\pi^{-}\pi^{-}\pi^{0}$      &253.0$\pm$0.6   & $24.54\pm0.06$  &7.83$\pm$0.17\\
      $D^-\to K^{0}_{S}\pi^{-}$                &93.3$\pm$0.3    & $50.90\pm0.17$ &9.99$\pm$0.33\\
      $D^-\to K^{0}_{S}\pi^{-}\pi^{0}$         &208.9$\pm$0.5   & $25.15\pm0.06$  &7.88$\pm$0.21\\
      $D^-\to K^{0}_{S}\pi^{-}\pi^{-}\pi^{+}$  &107.5$\pm$0.4   & $29.47\pm0.09$  &5.23$\pm$0.22\\
      $D^-\to K^{+}K^{-}\pi^{-}$               &68.9$\pm$0.3    & $41.23\pm0.18$  &9.00$\pm$0.35\\
      \bottomrule
    \end{tabular}
\end{table}

The one-dimensional fit projections to the $M^2_{\rm miss}$ and $M_{K_S^0\pi\pi}$ distributions of $D^0\to K_S^0\pi^-\pi^0e^+\nu_e$ and $D^+\to K_S^0\pi^+\pi^-e^+\nu_e$  are shown in Figure~\ref{combfit}. The fits return event yields of  $16.5^{+5.1}_{-4.5}$ for the  $D^0\to K_1(1270)^-e^+\nu_e$ signal and $20.2^{+6.2}_{-5.4}$ for the $D^+\to \bar{K}_1(1270)^0e^+\nu_e$ signal. The  DT yields and signal efficiencies are summarized in  Table~\ref{signal_eff}. The uncertainties in Table~\ref{signal_eff} are only statistical, and the systematic uncertainties will be discussed in Sec.~\ref{sys}.
The statistical significance of the signal  is estimated to be 5.4$\sigma$  for $D^0\to K_1(1270)^-e^+\nu_e$ and 5.6$\sigma$ for $D^+\to K_1(1270)^0e^+\nu_e$, by comparing the likelihoods with and without the signal component, and taking the change in the number of degrees of freedom into account. The significances of $D^0\to K_S^0\pi^-\pi^0 e^+\nu_e$ and $D^+\to K_S^0\pi^+ \pi^- e^+\nu_e$ are 4.5$\sigma$ and 4.0$\sigma$, respectively,  after accounting for systematic uncertainties associated with the 2D fits.

Inserting $N_{\rm DT}$, $\bar{\epsilon}_{\rm   sig}$, and $N_{\rm ST}^{\rm tot}$  into Eq.~\ref{equa} yields the BF for each decay.  Using the world average BFs of $\mathcal{B}({K_1(1270)^-\to K_S^0\pi^-\pi^0)}=(16.15\pm1.74)\%$ and $\mathcal{B}(\bar{K}_1(1270)^0\to K_S^0\pi^+$ $\pi^-)=(11.37\pm1.26)\%$\footnote{$\mathcal{B}_{K_1^0\to K_S^0\pi^+\pi^-}=\mathcal{B}_{K_1^0\to K^0\pi^+\pi^-} \times\frac{1}{2}\times \mathcal{B}_{K_S^0\to \pi^+\pi^-}=\frac{\mathcal{B}_{K_S^0\to \pi^+\pi^-}}{2}\times(\frac{1}{3}\times \mathcal{B}_{K_1\to K\rho}+\frac{4}{9}\times \mathcal{B}_{K_1\to K^*(892)\pi}+\frac{4}{9}\times \mathcal{B}_{K_1\to K_0^*(1430)\pi}\times \mathcal{B}_{K_0^*(1430)\to K\pi}$  $  + \mathcal{B}_{K_1\to K^+\omega}\times \mathcal{B}_{\omega\to \pi^+\pi^-}),~\mathcal{B}_{K_1^-\to K_S^0\pi^-\pi^0} = \mathcal{B}_{K_1^-\to K^0\pi^-\pi^0} \times\frac{1}{2}\times \mathcal{B}_{K_S^0\to \pi^+\pi^-}= \frac{\mathcal{B}_{K_S^0\to \pi^+\pi^-}}{2}\times(\frac{2}{3}\times \mathcal{B}_{K_1\to K\rho}+\frac{4}{9}\times \mathcal{B}_{K_1\to K^*(892)\pi}+\frac{4}{9}\times$ $  \mathcal{B}_{K_1\to K_0^*(1430)\pi}\times \mathcal{B}_{K_0^*(1430)\to K\pi}), \textup{where } K_1 \textup{ denotes } K_1(1270)  \textup{ and } K_S^0 \textup{ is  reconstructed via } K_S^0 \to \pi^+\pi^-$}, the absolute BFs of $D^0\to K_1(1270)^-e^+\nu_e$ and  $D^+\to \bar{K}_1(1270)^0 e^+\nu_e$ are also determined. 
The obtained BFs $\mathcal{B}_{\rm sig}$ are summarized in Table~\ref{tab:my_label}, and are in agreement with those obtained from measurements with a charged kaon in the final state~\cite{fylan,liuk}.  Therefore the two sets of the results are combined to yield  the values $\mathcal{B}_{\rm com}$, which are also given in Table~\ref{tab:my_label}.

\begin{table}[htb]
    \centering
    \caption{Summary of measured BFs $\mathcal{B}_{\rm sig}$ for different decays and combined  BFs $\mathcal{B}_{\rm com}$ for $D\to K_1(1270)e^+\nu_e$ decays. The first  and second  uncertainties are statistical and systematic, respectively.  For $D\to K_1(1270)e^+\nu_e$ modes, the third uncertainty originates from the assumed BFs of $K_1(1270)$ decays~\cite{pdg}.}
    
    \begin{tabular}{lcc}\hline
    \bottomrule

    Decay mode & $\mathcal{B}_{\rm sig}$ ($\times 10^{-4}$) & $\mathcal{B}_{\rm com}$ ($\times 10^{-4}$)\\ \hline
    $D^0\to K^0_S\pi^-\pi^0e^+\nu_e$ &  $(1.69^{+0.53}_{-0.46 }\pm0.18 )$       & / \\
    $D^+\to K^0_S\pi^+\pi^-e^+\nu_e$ & $(1.47^{+0.45}_{-0.40 }\pm0.20 )$  & / \\ 
    $D^0\to K_1(1270)^-e^+\nu_e$  & $(10.5^{+3.3}_{-2.8}\pm1.2\pm1.2)$   & $(10.8^{+1.3}_{-1.2} {}^{+0.8}_{-1.1}\pm1.2) $\\
    $D^+\to \bar K_1(1270)^0e^+\nu_e$ & $(12.9^{+4.0}_{-3.5}\pm1.8\pm1.5)$&  $(18.6^{+2.3}_{-2.2}\pm1.3\pm 2.1) $\\     \bottomrule
    \end{tabular}
    \label{tab:my_label}
\end{table}

\section{Systematic uncertainties}\label{sys}
 While in the BF determination using Eq.~\ref{equa} the uncertainties associated with the ST candidate selection cancel, the following sources of systematic uncertainties must be considered.

\begin{itemize}
    \item The uncertainty in the total yield  of ST $\bar{D}$ mesons is assigned to be 0.5\%~\cite{fylan,liuk}.
    \item $\pi^\pm$ tracking and PID efficiencies. 
    The data/MC differences of  $\pi^{\pm}$ tracking and PID efficiencies  are  re-weighted by the corresponding $\pi^{\pm}$ momentum spectra of signal MC events. The systematic uncertainty of tracking 
 (PID) efficiency is assigned to be $0.2\%$ (0.3\%) per $\pi^{\pm}$, based on the residual statistical uncertainties of the measured data/MC differences.
    
    \item $e^\pm$ tracking and PID efficiencies. The systematic uncertainties originating from $e^{+}$ tracking and PID efficiencies are studied by using a control sample of $e^{+} e^{-} \rightarrow \gamma e^{+} e^{-}$ events.  The  tracking and PID efficiencies of MC  are also re-weighted in momentum and $\cos \theta$ to match the $D \rightarrow K_S^0\pi\pi e^{+} \nu_e$ data. The systematic uncertainty of tracking (PID) efficiency is assigned to be $0.3\% $(0.3\%) per $e^{\pm}$.
    
    \item $K_S^0$ reconstruction. The systematic uncertainty associated with $K_S^0$ reconstruction is studied with control samples of the decays  $J/\psi\to K^{*\pm}K^{\mp}$ and $J/\psi\to \phi K_S^0K^{\pm}\pi^{\mp}$~\cite{KsSys}. The systematic uncertainty for each $K_S^0$ is assigned as 1.5\%.
    
    \item $\pi^0$ reconstruction. The systematic uncertainty of  $\pi^0$ reconstruction is assigned as 2.0\% per $\pi^0$ from studies of the DT $D^0 \bar{D}^0$ hadronic decay samples~\cite{BESIII:2016gbw}.
    
    \item Two-dimensional fit. To estimate the uncertainty arising from the signal shape used in the fit, the mass and width of $K_1(1270)$ are varied by $\pm1\sigma$. To take into account the potential resolution difference between data and MC simulation in the fit, a convolution of a Gaussian function is considered for $M_{\rm miss}^2$ and $M_{K_S^0\pi\pi}$. The peaking background yields are varied by $20.0\%$  after considering their statistical fluctuations. The uncertainties of combinatorial background shapes are estimated by varying  the smoothing parameters~\cite{roofit}. Potential  non-resonant  decays are  also considered and the uncertainty associated with this component is assigned as the change of the fitted DT yield by fixing its contribution~(14.2\%) ~\cite{Belle2011,fylan} in the two-dimensional fit. The associated  systematic uncertainties  are summarized in  Table~\ref{sys_2d}. 

    \item Signal generator.   To estimate the systematic uncertainty associated with the signal generator,  alternative signal MC events are generated using a phase-space model. The changes in the measured BFs using this alternative MC simulation are  6.5\% for $\mathcal{B}(D^0\to K_S^0\pi^-\pi^0e^+\nu_e)$ and 4.0\% for $\mathcal{B}(D^+ \to K_S^0 \pi^+\pi^-e^+\nu_e)$. 

    \item $K_1(1270)$ subdecays.  The uncertainties in the ratios of $K_1(1270)$ subdecays are  assigned by remeasuring the BFs based on PDG models~\cite{pdg}. A systematic uncertainty of 2.0\% is assigned both for $\mathcal{B}(D^0\to K_S^0\pi^-\pi^0e^+\nu_e)$ and $\mathcal{B}(D^+ \to K_S^0 \pi^+\pi^-e^+\nu_e)$.
    
    \item MC sample size. The systematic uncertainty due to the limited size of the MC sample is assigned to be 1.0$\%$  by $\sqrt{\sum_i({f_i\frac{\sigma_{\epsilon_i}}{\epsilon_{i}})^2}}$, where $f_i$ is the tag yield fraction, and $\epsilon_i$ and $\sigma_{\epsilon_i}$ are signal efficiency and the corresponding uncertainty of tag mode $i$, respectively.
    \item FSR  recovery. The uncertainty from FSR recovery is assigned to be 0.3\%  following  Ref.~\cite{BESIII:2015tql}.
\end{itemize}

\begin{table}[htbp]
    \centering
    \caption{Systematic uncertainties from the 2D fits.}
    \begin{tabular}{lcc}
    \hline
    \bottomrule
    Uncertainty~(\%) & $K_S^0\pi^-\pi^0e^+\nu_e$ & $K_S^0\pi^+\pi^-e^+\nu_e$ \\ 
    \hline
    Signal shape & 3.0 & 5.0\\
    Resolution& 2.4 & 3.3 \\
    Peaking background & 1.3 & 2.1  \\
    Combinatorial background& 1.0 & 3.8\\
    Non-resonant decays& 7.0 & 10.7 \\
\hline
Total & 8.2& 13.0\\
    \bottomrule

    \end{tabular}
    \label{sys_2d}
\end{table}

\begin{table}[htbp]
    \caption{Relative systematic uncertainties in the BF measurements.}
    \centering
    \begin{tabular}{l c c}
    \hline
    \bottomrule
         Uncertainty~(\%)&$K_S^0\pi^-\pi^0e^+\nu_e$&$K_S^0\pi^+\pi^-e^+\nu_e$   \\
         \hline
         $N_{\rm ST}$ & 0.5&0.5\\
         $\pi^\pm, e^\pm$ tracking & 0.5& 0.7\\
         $\pi^\pm, e^\pm$ PID & 0.6& 0.9\\
         $K_S^0$ reconstruction & 1.5&1.5\\
         $\pi^0$ reconstruction& 2.0&$/$\\
         Two-dimensional simultaneous fit & 8.2&13.0\\
         Signal generator& 6.5&4.0\\
         $K_1(1270)$ subdecays &2.0&2.0 \\
         MC sample size & 1.0 &1.0\\
         FSR recovery & 0.3&0.3\\
         \hline
         Total & 11.0 &14.0\\
         \bottomrule\\
         \label{uncertainties}
    \end{tabular}
\end{table}

The total systematic uncertainty is estimated  by adding all the individual contributions in quadrature. The sources of the systematic uncertainties in the BF measurements are summarized in Table~\ref{uncertainties}. They are assigned  relatively to the measured BFs.

%------------------------------------------------------------------------------
\section{Summary}

By analyzing a data sample corresponding to an integrated luminosity  of 2.93 $\rm fb^{-1}$ collected at $\sqrt s=3.773$ GeV with the BESIII detector, the first observations of the semileptonic decays $D^0\to K_S^0\pi^-\pi^0e^+\nu_e$ and $D^+\to K_S^0\pi^+\pi^-e^+\nu_e$ are obtained with statistical significances of $5.3\sigma$ and $5.6\sigma$, respectively. The resulting BFs are summarized in Table~\ref{tab:my_label}. The measured BFs of $D\to \bar{K}_1(1270)e^+\nu_e$ decays are consistent with previous measurements using $\bar{K}_1(1270)\to K^-\pi^+\pi^{(-,0)}$~\cite{fylan,liuk}. The combined BFs of $D\to \bar{K}_1(1270)e^+\nu_e$ agree with the CLFQM and LCSR predictions when $\theta_{K_1}\approx$  33° or 57° \cite{Cheng:2017pcq,Momeni:2020zrb} and contradict the predictions reported in Ref. \cite{Momeni:2019uag} when setting the value of $\theta_{K_{1}}$ negative.

With approximately six times more data coming from BESIII at $\sqrt{s} = 3.773\;\text{GeV}$ in the foreseen future~\cite{BESIII:2020nme}, a thorough investigation with the enlarged data samples in the four $K_1$ channels $\left(K^{-} \pi^{+} \pi^{+}, K_S^0 \pi^0 \pi^{+}, K_S^0 \pi^{+} \pi^{-}, K^{-} \pi^{+} \pi^0\right)$ will be possible to further elucidate the knowledge on $K_1(1270)$ and $K_1(1400)$ meson in a systematic fashion.

% \end{linenumbers}
\acknowledgments
The BESIII Collaboration thanks the staff of BEPCII and the IHEP computing center for their strong support. This work is supported in part by National Key R\&D Program of China under Contracts Nos. 2020YFA0406400, 2020YFA0406300; National Natural Science Foundation of China (NSFC) under Contracts Nos. 11635010, 11735014, 11835012, 11935015, 11935016, 11935018, 11961141012, 12022510, 12025502, 12035009, 12035013, 12061131003, 12192260, 12192261, 12192262, 12192263, 12192264, 12192265, 12221005, 12225509, 12235017; the Chinese Academy of Sciences (CAS) Large-Scale Scientific Facility Program; the CAS Center for Excellence in Particle Physics (CCEPP); Joint Large-Scale Scientific Facility Funds of the NSFC and CAS under Contract No. U1832207; CAS Key Research Program of Frontier Sciences under Contracts Nos. QYZDJ-SSW-SLH003, QYZDJ-SSW-SLH040; 100 Talents Program of CAS; The Institute of Nuclear and Particle Physics (INPAC) and Shanghai Key Laboratory for Particle Physics and Cosmology; ERC under Contract No. 758462; European Union's Horizon 2020 research and innovation programme under Marie Sklodowska-Curie grant agreement under Contract No. 894790; German Research Foundation DFG under Contracts Nos. 443159800, 455635585, Collaborative Research Center CRC 1044, FOR5327, GRK 2149; Istituto Nazionale di Fisica Nucleare, Italy; Ministry of Development of Turkey under Contract No. DPT2006K-120470; National Research Foundation of Korea under Contract No. NRF-2022R1A2C1092335; National Science and Technology fund of Mongolia; National Science Research and Innovation Fund (NSRF) via the Program Management Unit for Human Resources \& Institutional Development, Research and Innovation of Thailand under Contract No. B16F640076; Polish National Science Centre under Contract No. 2019/35/O/ST2/02907; The Swedish Research Council; U. S. Department of Energy under Contract No. DE-FG02-05ER41374.

\clearpage
%\author{Author list}
%\begin{small}
%\begin{center}
\large
The BESIII Collaboration\\
\normalsize
%% Saved at => 2023-03-31
\\ M.~Ablikim$^{1}$, M.~N.~Achasov$^{5,b}$, P.~Adlarson$^{75}$, X.~C.~Ai$^{81}$, R.~Aliberti$^{36}$, A.~Amoroso$^{74A,74C}$, M.~R.~An$^{40}$, Q.~An$^{71,58}$, Y.~Bai$^{57}$, O.~Bakina$^{37}$
% ...
\\
\vspace{0.2cm}
(BESIII Collaboration)\\
\vspace{0.2cm} {\it
$^{1}$ Institute of High Energy Physics, Beijing 100049, People's Republic of China\\
% ...
$^{5}$ Budker Institute of Nuclear Physics SB RAS (BINP), Novosibirsk 630090, Russia\\
% ...
$^{36}$ Johannes Gutenberg University of Mainz, Johann-Joachim-Becher-Weg 45, D-55099 Mainz, Germany\\
$^{37}$ Joint Institute for Nuclear Research, 141980 Dubna, Moscow region, Russia\\
% ...
$^{40}$ Liaoning Normal University, Dalian 116029, People's Republic of China\\
% ...
$^{57}$ Southeast University, Nanjing 211100, People's Republic of China\\
$^{58}$ State Key Laboratory of Particle Detection and Electronics, Beijing 100049, Hefei 230026, People's Republic of China\\
% ...
$^{71}$ University of Science and Technology of China, Hefei 230026, People's Republic of China\\
% ...
$^{74}$ University of Turin and INFN, (A)University of Turin, I-10125, Turin, Italy; (B)University of Eastern Piedmont, I-15121, Alessandria, Italy; (C)INFN, I-10125, Turin, Italy\\
$^{75}$ Uppsala University, Box 516, SE-75120 Uppsala, Sweden\\
% ...
$^{81}$ Zhengzhou University, Zhengzhou 450001, People's Republic of China

\vspace{0.2cm}
\noindent $^{a}$ Also at the Moscow Institute of Physics and Technology, Moscow 141700, Russia\\
$^{b}$ Also at the Novosibirsk State University, Novosibirsk, 630090, Russia\\
% ...
}
%% ends here %%
%\end{center}
%\vspace{0.4cm}
%\end{small}


\begin{thebibliography}{99}

\bibitem{lepdecay}
N.~Isgur, D.~Scora, B.~Grinstein and M.~B.~Wise,
\textit{Semileptonic B and D decays in the quark model},
\href{doi:10.1103/PhysRevD.39.799}{~Phys. Rev. D \textbf{39} (1989), 799}.

% \cite{Scora:1995ty}
\bibitem{isgw2}
D.~Scora and N.~Isgur,
\textit{Semileptonic meson decays in the quark model: an update},\href{doi:10.1103/PhysRevD.52.2783}{~Phys. Rev. D \textbf{52} (1995), 27832812}.

\bibitem{mixing}H.~Hatanaka and K.~C.~Yang,
\textit{$K_1(1270)-K_1(1400)$ Mixing angle and new-physics effects in $B\to K_1\ell^+\ell^-$ decays},\href{doi:10.1103/PhysRevD.78.074007}{~Phys. Rev. D \textbf{78} (2008), 074007}.

\bibitem{Cheng:2017pcq}
H.~Y.~Cheng and X.~W.~Kang, \textit{Branching fractions of semileptonic $D$ and $D_s$ decays from the covariant light-front quark model}, \href{doi:10.1103/PhysRevD.92.072012}{Eur. Phys. J. C \textbf{77} (2017) no.9, 587}.

\bibitem{Momeni:2020zrb}
S.~Momeni, \textit{Helicity form factors for $D_{(s)} \rightarrow A \ell \nu $ process in the light-cone QCD sum rules approach}, \href{doi:10.1103/PhysRevD.92.072012}{Eur. Phys. J. C \textbf{80} (2020) no.6, 553}.

\bibitem{Momeni:2019uag}
S.~Momeni and R.~Khosravi, \textit{Semileptonic $D_{(s)} \to A \ell^+ \nu$ and nonleptonic $D\to K_1(1270,1400) \pi$} \textit{ decays in LCSR}, \href{https://iopscience.iop.org/article/10.1088/1361-6471/ab35d0}{J. Phys. G \textbf{46} (2019) no.10, 105006}.

\bibitem{pdg} Particle Data Group,
\textit{Review of Particle Physics}, \href{doi:10.1093/ptep/ptac097}{PTEP \textbf{2022} (2022), 083C01}.


\bibitem{Wang:2019wee} W.~Wang, F.~S.~Yu and Z.~X.~Zhao, \textit{Novel method to reliably determine the photon helicity in  $B\to K_{1}\gamma$},\href{doi:10.1103/PhysRevLett.125.051802}{~Phys. Rev. Lett. \textbf{125} (2020), 051802}.

\bibitem{Bian:2021gwf}
L.~Bian, L.~Sun and W.~Wang, \textit{Up-down asymmetries and angular distributions in $D\to K_1\ell^+\nu_\ell$},\href{doi:10.1103/PhysRevD.104.053003}{~Phys. Rev. D \textbf{104} (2021), 053003}.


\bibitem{liuk} BESIII Collaboration,
\textit{Observation of the semileptonic $D^+$ decay into the $\bar K_1(1270)^0$ axial-vector meson},\href{doi:10.1103/PhysRevLett.123.231801}{~Phys. Rev. Lett. \textbf{123} (2019), 231801}.

\bibitem{fylan} BESIII Collaboration,
\textit{Observation of $D^0\to K_1(1270)^- e^+\nu_e$},\href{doi:10.1103/PhysRevLett.127.131801}{~Phys. Rev. Lett. \textbf{127} (2021), 131801}.


\bibitem{Belle2011} Belle Collaboration,
\textit{Study of the $K^+ \pi^+ \pi^-$ final state in $B^+ \to J/\psi K^+ \pi^+ \pi^-$ and $B^+ \to \psi^\prime K^+ \pi^+ \pi^-$},\href{doi:10.1103/PhysRevD.83.032005}{~Phys. Rev. D \textbf{83} (2011), 032005}.



\bibitem{dskkpipi0}
BESIII Collaboration, \textit{Amplitude analysis and branching fraction measurement of $D_s^+ \to K^-K^+\pi^+\pi^0$},\href{doi:10.1103/PhysRevD.104.032011}{~Phys. Rev. D \textbf{104} (2021), 032011}.

\bibitem{2012D0kkpipi1}CLEO Collaboration,
\textit{Amplitude analysis of $D^0\to K^+K^-\pi^+\pi^-$},\href{doi:10.1103/PhysRevD.85.122002}{~Phys. Rev. D \textbf{85} (2012), 122002}.

\bibitem{kpipipi} BESIII Collaboration,
\textit{Amplitude analysis of $D^{0} \rightarrow K^{-} \pi^{+} \pi^{+} \pi^{-}$},\href{doi:10.1103/PhysRevD.95.072010}{~Phys. Rev. D \textbf{95} (2017), 072010}.


%\cite{LHCb:2017swu}
\bibitem{lhcb17}LHCb Collaboration, \textit{Studies of the resonance structure in $D^{0} \rightarrow K^\mp \pi ^\pm \pi ^\pm \pi ^\mp $ decays},\href{doi:10.1140/epjc/s10052-018-5758-4}{~Eur. Phys. J. C \textbf{78} (2018), 443}.

%\cite{dArgent:2017gzv}
\bibitem{4pi}
 P. d'Argent {\it et al.},
\textit{Amplitude analyses of $D^0 \to {\pi^+\pi^-\pi^+\pi^-}$ and $D^0 \to {K^+K^-\pi^+\pi^-}$ Decays},\href{doi:10.1007/JHEP05(2017)143}{~JHEP \textbf{05} (2017), 143}.

%\cite{Guo:2018orw}
\bibitem{Ddecays}
P.~F.~Guo, D.~Wang and F.~S.~Yu,
\textit{Strange axial-vector mesons in $D$ meson decays},\href{doi:10.11804/NuclPhysRev.36.02.125}{~Nucl. Phys. Rev. \textbf{36} (2019), 125134}.
%\cite{Ablikim:2013ntc}

\bibitem{Ablikim:2013ntc}BESIII Collaboration,\textit{~Measurement of the integrated luminosities of the data taken by BESIII at $\sqrt{s}=$3.650 and 3.773 GeV},\href{doi:10.1088/1674-1137/37/12/123001}{~Chin. Phys. C \textbf{37} (2013), 123001}.

%\cite{BESIII:2009fln}
\bibitem{Ablikim:2009aa}
BESIII Collaboration,\textit{~Design and construction of the BESIII detector},\href{doi:10.1016/j.nima.2009.12.050}{~Nucl. Instrum. Meth. A \textbf{614} (2010), 345399}.

%\cite{Yu:2016cof}
\bibitem{BEPCII}
C. H. Yu, Y. Zhang, Q. Qin, J.Q. Wang and G. Xu, \textit{et al.}, \textit{BEPCII performance and beam dynamics studies on luminosity},\href{http://jacow.org/ipac2016/papers/tuya01.pdf}{~Proceedings of IPAC2016, Busan, Korea, 2016, doi:10.18429/JACoW-IPAC2016-TUYA01.}

%\cite{BESIII:2020nme}
\bibitem{BESIII:2020nme}
BESIII Collaboration,\textit{~Future physics programme of BESIII},\href{doi:10.1088/1674-1137/44/4/040001}{~Chin. Phys. C \textbf{44} (2020), 040001}.



%\cite{GEANT4:2002zbu}
\bibitem{geant4}GEANT4 Collaboration, \textit{GEANT4--a simulation toolkit},\href{doi:10.1016/S0168-9002(03)01368-8}{~Nucl. Instrum. Meth. A \textbf{506} (2003), 250303}.

\bibitem{detvis}
  K.~X.~Huang, {\it et al.},  \textit{Method for detector description transformation to Unity and application in BESIII}, \href{https://doi.org/10.1007/s41365-022-01133-8}{~Nucl.\ Sci.\ Tech. {\bf 33}, 142 (2022)}.


  
%\cite{Jadach:2000ir}
\bibitem{kkmc}S.~Jadach, B.~F.~L.~Ward and Z.~Was,
\textit{Coherent exclusive exponentiation for precision Monte Carlo calculations},\href{doi:10.1103/PhysRevD.63.113009}{~Phys. Rev. D \textbf{63} (2001), 113009}.

%\cite{Jadach:1999vf}
\bibitem{kkmc2}
S.~Jadach, B.~F.~L.~Ward and Z.~Was,\textit{~The precision Monte Carlo event generator KK for two fermion final states in $e^+ e^-$ collisions,}\href{doi:10.1016/S0010-4655(00)00048-5}{~Comput. Phys. Commun. \textbf{130} (2000), 260325}.



%\cite{Ping:2008zz}
\bibitem{evtgen}R.~G.~Ping,\textit{~Event generators at BESIII},\href{doi:10.1088/16741137/32/8/001}{~Chin. Phys. C \textbf{32} (2008), 599}.

%\cite{Lange:2001uf}
\bibitem{evtgen2}
D.~J.~Lange,\textit{~The EvtGen particle decay simulation package,}\href{doi:10.1016/S0168-9002(01)00089-4}{~Nucl. Instrum. Meth. A \textbf{462} (2001), 152155}.


%\cite{Chen:2000tv}
\bibitem{lundcharm}
J.~C.~Chen, G.~S.~Huang, X.~R.~Qi, D.~H.~Zhang and Y.~S.~Zhu,
\textit{Event generator for $J/\psi$ and $\psi(2S)$ decay},\href{doi:10.1103/PhysRevD.62.034003}{~Phys. Rev. D \textbf{62} (2000), 034003}.
\bibitem{lundcharm2}
R.~L.~Yang, R.~G.~Ping and H.~Chen,\textit{Tuning and validation of the Lundcharm model with $J/\psi$ decays},
\href{doi:10.1088/0256-307X/31/6/061301}{Chin. Phys. Lett. \textbf{31} (2014), 061301}.


%\cite{Richter-Was:1992hxq}
\bibitem{photos}
E.~Richter-Was,
\textit{QED bremsstrahlung in semileptonic B and leptonic tau decays},\href{doi:10.1016/0370-2693(93)90062-M}{Phys. Lett. B \textbf{303} (1993), 163}.

\bibitem{MARK-III:1985hbd}
MARK-III  Collaboration,
% R.~M.~Baltrusaitis \textit{et al.} [MARK-III],
%``Direct Measurements of Charmed d Meson Hadronic Branching Fractions,''
\textit{Direct Measurements of Charmed $D$ Meson Hadronic Branching Fractions}, \href{https://journals.aps.org/prl/abstract/10.1103/PhysRevLett.56.2140}{Phys. Rev. Lett. \textbf{56} (1986), 2140}.


\bibitem{Li:2021iwf}
H.~B.~Li and X.~R.~Lyu,
\textit{Study of the standard model with weak decays of charmed hadrons at BESIII},
\href{https://inspirehep.net/files/55ea2734d743c42aeaac0780c70dd4fb}{~Natl. Sci. Rev. \textbf{8} (2021), nwab181}.





%\cite{BESIII:2016gbw}
\bibitem{BESIII:2016gbw}BESIII Collaboration,\textit{~Improved measurement of the absolute branching fraction of $D^{+}\to\bar{K}^0 \mu ^{+}\nu _{\mu }$},\href{doi:10.1140/epjc/s10052-016-4198-2}{~Eur. Phys. J. C \textbf{76} (2016), 369}.

\bibitem{k0enu}BESIII Collaboration,
\textit{Measurement of the absolute branching fraction of $D^{+}\rightarrow\bar K^0 e^{+}\nu_{e}$ via $\bar K^0\to\pi^0\pi^0$},\href{doi:10.1088/1674-1137/40/11/113001}{~Chin. Phys. C \textbf{40} (2016, 113001}.


%\cite{BESIII:2018nzb}
\bibitem{pimunu}BESIII Collaboration,
\textit{Measurement of the branching fraction for the semi-leptonic decay $D^{0(+)}\to \pi^{-(0)}\mu^+\nu_\mu$ and test of lepton universality},\href{doi:10.1103/PhysRevLett.121.171803}{Phys. Rev. Lett. \textbf{121} (2018), 171803}.

%\cite{BESIII:2018sjg}
\bibitem{a0enu}BESIII Collaboration,\textit{~Observation of the semileptonic decay $D^0 \to a_0(980)^- e^+ \nu_e$ and evidence for $D^+ \to a_0(980)^0 e^+ \nu_e$},\href{doi:10.1103/PhysRevLett.121.081802}{~Phys. Rev. Lett. \textbf{121} (2018), 081802}.

%\cite{BESIII:2018ccy}
\bibitem{kmunu}BESIII Collaboration,\textit{~Study of the $D^0\to K^-\mu^+\nu_\mu$ dynamics and test of lepton flavor universality with $D^0\to K^-\ell^+\nu_\ell$ decays},\href{doi:10.1103/PhysRevLett.122.011804}{~Phys. Rev. Lett. \textbf{122} (2019), 011804}.

%\cite{BESIII:2018qmf}
\bibitem{f0enu}BESIII Collaboration,\textit{~Observation of $D^+ \to f_0(500) e^+\nu_e$ and improved measurements of $D \to\rho e^+\nu_e$},\href{doi:10.1103/PhysRevLett.122.062001}{~Phys. Rev. Lett. \textbf{122} (2019), 062001}.

%\cite{BESIII:2014rtm}
\bibitem{kpicosmic}
BESIII Collaboration,\textit{~Measurement of the $D\to K^-\pi^+$ strong phase difference in $\psi(3770)\to D^0\overline{D}{}^0$},\href{doi:10.1016/j.physletb.2014.05.071}{~Phys. Lett. B \textbf{734} (2014), 227233}.

%\cite{ARGUS:1990hfq}
\bibitem{ARGUS}ARGUS Collaboration,\textit{~Search for hadronic $b \to u$ decays},\href{doi:10.1016/0370-2693(90)91293-K}{~Phys. Lett. B \textbf{241} (1990), 278282}.




%\cite{Verkerke:2003ir}
\bibitem{roofit}W.~Verkerke and D.~P.~Kirkby,\textit{~The RooFit toolkit for data modeling},\href{[arXiv:physics/0306116 [physics]]}{~eConf \textbf{C0303241} (2003), MOLT007}.

%\cite{BESIII:2018jjm}
\bibitem{KsSys}
M.~Ablikim \textit{et al.}BESIII Collaboration, \textit{Study of the decay $D^0\rightarrow \bar{K}^0\pi^-e^+\nu_e$}, \href{doi:10.1103/PhysRevD.99.011103}{Phys. Rev. D \textbf{99} (2019), 011103}.


%\cite{BESIII:2015tql}
\bibitem{BESIII:2015tql}BESIII Collaboration,\textit{Study of dynamics of $D^0 \to K^- e^+ \nu_{e}$ and $D^0\to\pi^- e^+ \nu_{e}$ decays},\href{doi:10.1103/PhysRevD.92.072012}{~Phys. Rev. D \textbf{92} (2015) no.7, 072012}.





\end{thebibliography}
\end{document}